\documentclass[proof]{WileyASNA-v1}

\articletype{Article Type}%

\received{}
\revised{}
\accepted{}

\begin{document}

\title{Hidden No More: Spotlight on tidal disruption events in active galactic nuclei}

\author[1]{Patrik Mil\'an Veres*}

\authormark{Patrik Mil\'an Veres \textsc{et al}}

\address[1]{
  \orgdiv{Faculty of Physics and Astronomy, Astronomical Institute (AIRUB)},
  \orgname{Ruhr University Bochum},
  \orgaddress{
    \street{Universitätsstraße 150},
    \postcode{44801},
    \city{Bochum},
    \country{Germany}
  }
}

\corres{*\email{veres@astro.ruhr-uni-bochum.de}}

\abstract{Tidal disruption events (TDEs) are typically discovered in previously quiescent galaxies. However, earlier studies have revealed a handful of TDEs occurring in pre-existing active galactic nuclei (AGNs). We discuss AT2019aalc, a promising TDE candidate in an AGN, and compare it to similar sources. We also explore Bowen fluorescence flares, a newly identified class of flaring supermassive black holes, as potential members of the TDE in AGN transient class. We aim to connect the observed properties of these flares with the expectations of TDE-in-AGN simulations.}



\maketitle

\section{Introduction}\label{sec1}
Nuclear transients are of significant scientific interest. Especially, tidal disruption events (TDEs) which are rare transients occurring when a star approaches the vicinity of a supermassive black hole (SMBH) and the tidal forces of the latter rip the star apart \citep{1988Natur.333..523R}. TDEs provide a unique opportunity to study SMBHs in otherwise non-active galactic nuclei, where such black holes are typically difficult to detect. However, TDEs are expected to occur not only in previously quiescent galaxies but also in active galactic nuclei (AGN). Although, their event rates should be comparable or even higher \citep{2025ApJ...979..172K} than that of regular TDEs in quiescent galaxies, the number of known TDEs in AGNs (hereafter TDE-AGNs) remains very small, mainly because distinguishing them from other forms of AGN variability is challenging. Identified promising candidates include PS16dtm \citep{ps16dtm}, AT2021aeuk \citep{at2021aeuk_jingbo}, AT2022agi (in ULIRG F01004-2237) \citep{f01,f01_partial}, AT2020zso \citep{Wevers22} and AT2019aalc \citep{lancel}. Studying TDE-AGN can refine estimates of TDE rates and may help explain episodes of AGN activity that deviate from their regular behavior. Moreover, TDE-AGNs are expected to exhibit luminous optical flares and therefore potentially provide crucial insights into repeating partial TDEs, which produce fainter outbursts than complete disruptions. Repeating partial tidal disruption events (pTDEs) can be envisioned as a natural outcome of the Hills capture mechanism \citep{Hills88}, leading to episodic stripping and recurring flares on timescales of months to years.

Several classes of nuclear transients with uncertain physical origin may in fact include TDE-AGNs, such as some of the extreme nuclear transients (ENTs; e.g., \cite{2025NatAs.tmp..214G}) or changing-look AGN \citep[e.g.,][]{2024ApJ...962L...7W}. \cite{Frederick20} suggested that some of the transients primarily associated with flaring narrow-line Seyfert-1 galaxies are possibly powered by TDEs. However, a unified and comprehensive classification of TDE-AGNs is still missing. Simulations of TDE-AGN suggest that these events should differ from regular TDEs; their SED may not be purely thermal due to the underlying AGN continuum \citep{tdes_in_agn}. Furthermore, interactions between the pre-existing accretion disk and the stellar debris may significantly influence the overall flux, primarily in optical/UV and X-rays \citep{tdes_in_agn,2024MNRAS.527.8103R}. 

In this work, we review the observed properties of the nuclear transient AT2019aalc (Section \ref{sec2}) and compare its key properties to similar transients (Section \ref{sec3}). We elaborate on whether BFFs represent a previously hidden population of TDEs occurring in previously active galaxies (Section \ref{sec4}).

\section{AT2019aalc: When AGN Meets TDE?}\label{sec2}
The nuclear transient AT2019aalc is hosted by a broad-line Seyfert-1 galaxy and exhibited two distinct luminous optical flares. The transient was not at the focus of attention during its initial flare in 2019. Later, an archival search implied an association with a high-energy neutrino detected by the IceCube Observatory \citep{neu} making AT2019aalc as a potentially unique transient. Furthermore, a second flare occurred in 2023, roughly four years after the first flare.

Some of the multi-wavelength properties observed during the second optical flare of AT2019aalc are reminiscent to TDEs. These are the UV-bright nature of the transient, very soft X-ray emission, delayed radio flare (with respect to the optical flare) and the IR dust echo. Moreover, no significant optical variability was observed prior the optical flares and the host galaxy is not presented in any X-ray catalogs before 2023. This suggests a sudden enhancement of accretion. The double flare is consistent with a repeating pTDE scenario (see more detail in \cite{lancel}). On the other hand, the slowly declining optical emission (different from the canonical $t^{-5/3}$ fallback rate) and the cooling observed during this phase, are not commonly observed in TDEs. 

Spectroscopically, AT2019aalc exhibited with unique features that do not characterize regular AGN activity. Especially, the line features around 4660~\AA, which, in fact is known as the Bowen fluorescence line complex. This is due to a line fluorescence process that excites O\,{\sc iii} and N\,{\sc iii} lines in the optical and near-UV \citep{1934PASP...46..146B,1935ApJ....81....1B}. TDEs, which exhibit strong UV radiation and a dense photosphere suitable for the Bowen mechanism, generally show Bowen lines. However, similar to its photometric properties, AT2019aalc appeared to be a potentially peculiar case. Namely, although the variable Bowen lines and the blue continuum strongly suggest a TDE case, AT2019aalc exhibited with clearly narrower Balmer lines than those observed in TDEs (which typically characterized by a broad component of $> 10^{4}$\,km s$^{-1}$). AT2019aalc also displayed strong and variable high-ionization coronal lines (e.g., [Fe\,{\sc x}]$\lambda 6375$) which are, once again, indicative of a TDE. 

\section{AT2019aalc and Its Analogues}\label{sec3}
AT2019aalc exhibited several multi-wavelength properties typical of TDEs, yet also showed notable differences. AT2019aalc is not unique in this regard, as other transients with similar characteristics have been observed. These are identified as Bowen fluorescence flares (BFFs), known as a new class of accreting SMBHs \citep{Trakhtenbrot19a}. AT2019aalc shares multi-wavelength photometric and spectroscopic properties with these transients which allow to classify it as a BFF, in nice agreement with the interpretation by \cite{Marzena}. AT2019aalc is also as an extreme coronal line emitter (ECLE). ECLEs are thought to be associated with TDEs occurring in dusty environments \citep[e.g.,][]{Wang12}.

\begin{figure}[!htbp]
    \centering
    \includegraphics[width=\linewidth]{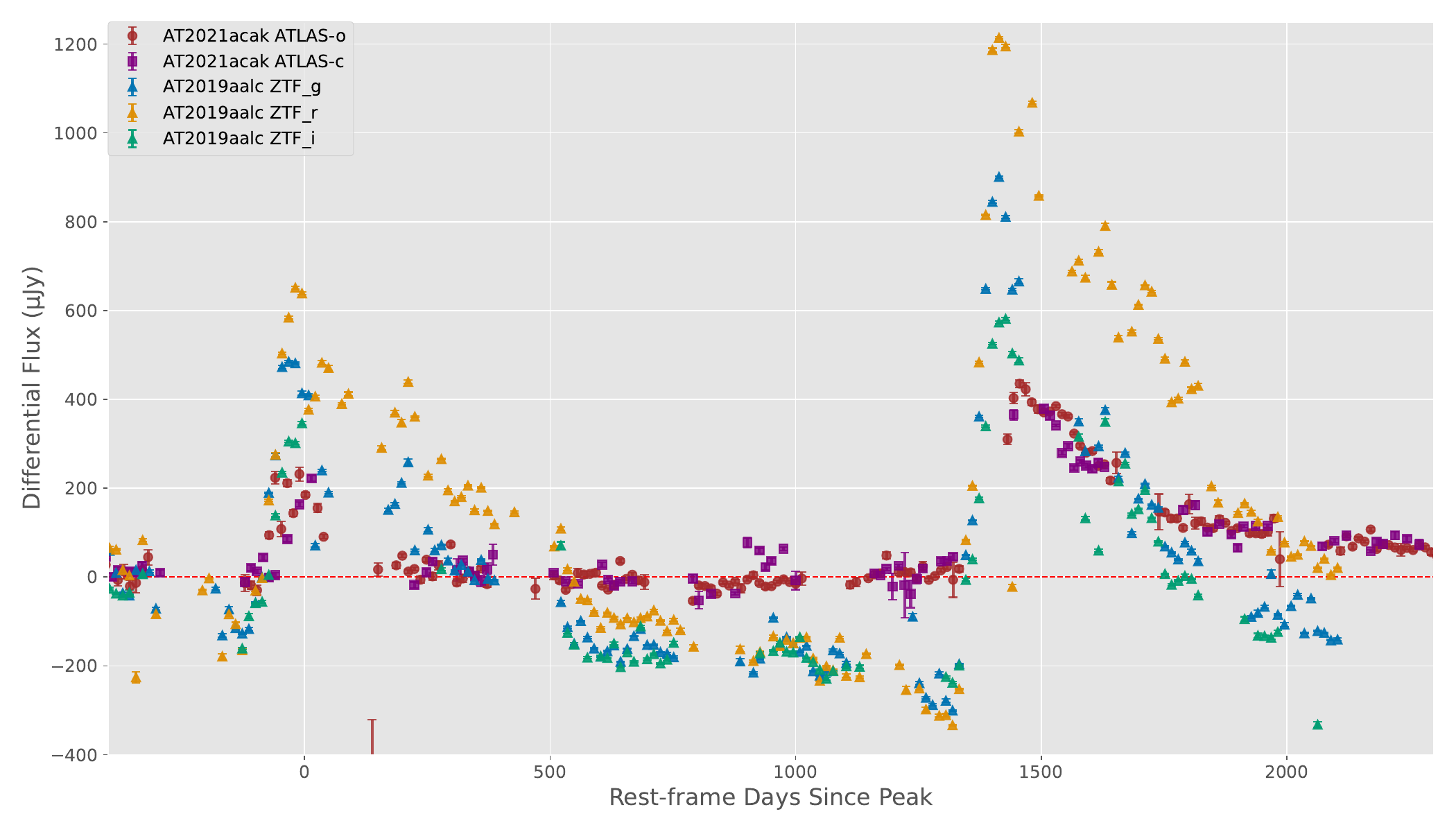}
    \caption{ZTF and ATLAS differential light curves of AT2019aalc and AT2021acak, respectively, retrieved via forced photometry requests and binned to $14$-days. The red dotted line indicates the baseline flux level. The days are relative to the first flare's peak. The similarities of the optical light curve properties are remarkable. AT2019aalc is known as a BFF while AT2021acak is a candidate TDE. \label{fig1}}
\end{figure}

In literature, we found the candidate TDE AT2021acak exhibiting with similar optical light curve properties to AT2019aalc; this nuclear transient also showed two distinct optical flares, with the second flare being more luminous and separated from the first by approximately four years. We show the extinction-corrected \citep{2011ApJ...737..103S} Zwicky Transient Facility (ZTF)\footnote{\url{https://ztfweb.ipac.caltech.edu/cgi-bin/requestForcedPhotometry.cgi}} \citep{2019PASP..131a8002B} and Asteroid Terrestrial-impact Last Alert System   
(ATLAS)\footnote{\url{https://fallingstar-data.com/forcedphot}} \citep{2021TNSAN...7....1S} light curves of AT2019aalc and AT2021acak, respectively, in Figure \ref{fig1}. Furthermore, an investigation of the optical spectroscopic properties of AT2021acak reveals additional similarities: the N{\sc iii} $\lambda 4640$ + He{\sc ii} $\lambda 4686$ Bowen fluorescence complex is prominently present, and several high-ionization lines are also detected, notably exhibiting measurable flux evolution \citep{acak}. The FWHM of the broad component of Balmer lines ($\approx 3000$\,km s$^{-1}$) \citep{acak} is clearly lower than typically measured in TDEs but very similar to the measured value for AT2019aalc. Both optical flares were accompanied by infrared dust echo flares, providing an additional point of similarity. We propose that AT2021acak belongs to the BFF class, representing another example of such a transient exhibiting two distinct optical flares, similar to AT2019aalc and ULIRG F01004-2237.

\section{Bowen flares in a broader framework}\label{sec4}
\begin{figure}[!htbp]
    \centering
    \includegraphics[width=0.88\linewidth]{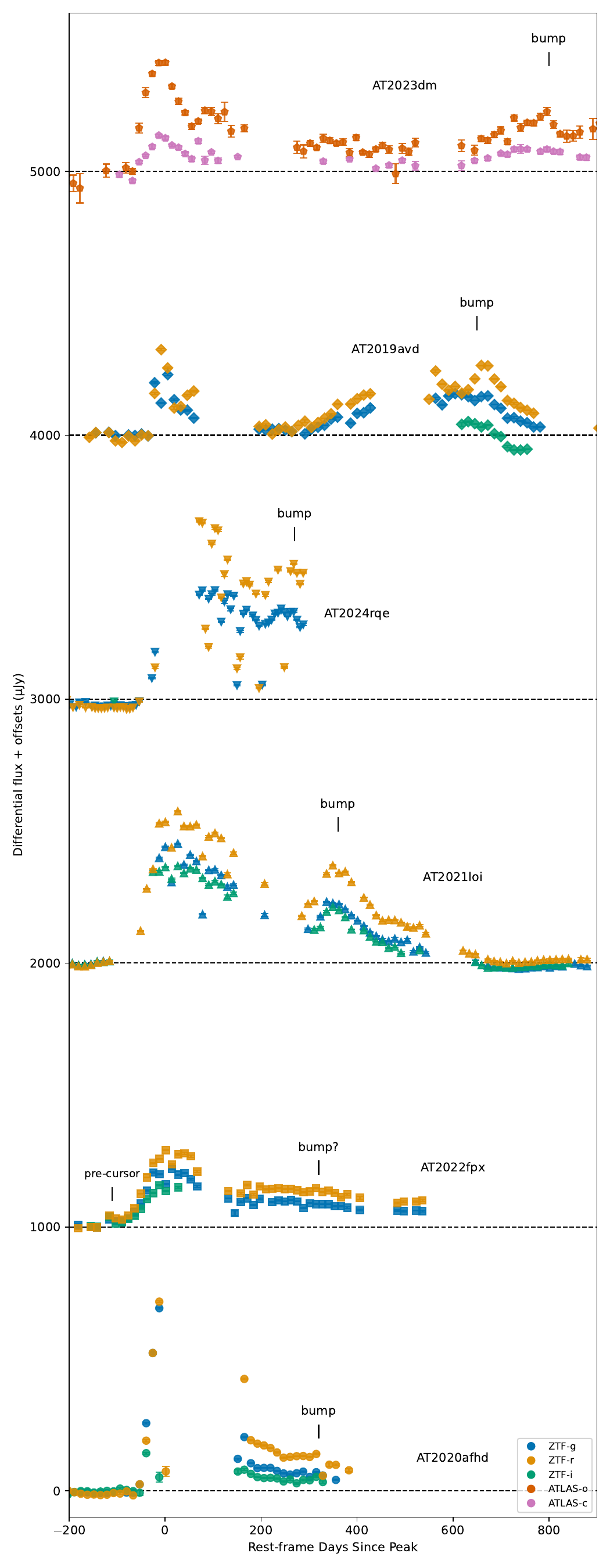}
    \caption{ZTF and ATLAS light curves of BFFs and BFF-like transients which show a late-time bump-like structure in their optical light curves. This bump feature differ from the distinct flares seen in a few other BFFs including AT2019aalc. The light curves are binned to $14$\,days, except for AT2024rqe, which is binned to $7$\,days. The black dashed lines represent the baseline flux levels for the different cases. Vertical offsets have been applied to all sources except AT2020afhd for clarity. A scaling factor of $1/2$ has been applied to AT2024rqe and AT2023dm to improve visibility. The peak of AT2024rqe and AT2020afhd are approximate due to lack of observations around that time. \label{fig2}}
\end{figure}

The first three BFFs (AT2017bgt, OGLE17aaj and ULIRG F01004-2237) were discovered by \cite{Trakhtenbrot19a}. Since then, only a handful of additional BFFs have been revealed. These are AT2021loi \cite{2021loi}, AT2019aalc \cite{lancel,Marzena}, and J012026 \cite{j01}. Additionally, a few more transients reminiscent to BFFs have been identified: AT2022fpx \citep{2024MNRAS.532..112K}, AT2020afhd \citep{2024TNSAN..53....1A}, AT2019avd \citep{2020TNSAN.105....1T}, AT2024rqe \citep{2025TNSCR.378....1P} and AT2022wtn \citep{2025MNRAS.540..498O}. In an archival search, we revealed AT2023dm as a BFF candidate. Three instances, AT2019aalc, AT2021acak, and ULIRG F01004-2237, discussed above, displayed two distinct optical flares. The two flares of AT2019aalc evolved very similarly (see Figure 6. in \citep{2025arXiv250706296J}). We also note that in fact almost none of the known BFFs exhibited a smooth decline. \cite{2021loi} and \cite{lancel} pointed out bump features in AT2017bgt, OGLE17aaj, ULIRG F01004-2237 and AT2021loi. In Figure \ref{fig2}., we display the extinction-corrected \citep{2011ApJ...737..103S} ZTF and ATLAS differential light curves of other BFFs and BFF candidates which exhibited remarkable bumps in their optical light curves. We note that these bumps differ from the three cases with re-brightening flares discussed earlier. These bump features either appear during the declining phase of the optical light curve or, if they appear later -- as in AT2019avd, more than a year afterward -- they exhibit a more symmetric shape distinct from the original flare. Furthermore, in addition to the re-brightening and bump features, a third feature of some BFF light curves should be highlighted. AT2019aalc exhibited quasi-periodic oscillations (QPOs) in the optical, as recurrent small-amplitude bumps during the decline of both main optical flares. Optical polarization observations reveal that the polarization angle changes alongside these QPOs \citep{2025arXiv250706296J}.

Importantly, at least some of the BFFs, for example AT2017bgt \citep{Trakhtenbrot19a}, AT2019aalc \citep{lancel} and AT2021loi \citep{2021loi}, took evidently place in previously-known AGNs. These active galaxies had been in a state of low-level activity for (at least) several years prior the observed transients. The sudden increase in optical/UV emission is a common characteristic of known BFFs, although AT2022fpx exhibited an optical precursor \citep{2024MNRAS.532..112K} a few months before the primary peak.

The multi-wavelength properties of BFFs exhibit considerable diversity. While some, such as AT2019aalc \citep{lancel,Marzena}, AT2019avd \citep{2022ApJ...928...63C}, and AT2020afhd \citep{2025arXiv251112477W}, are characterized by a strong soft X-ray excess, others, including AT2021loi \citep{2021loi} and AT2024rqe, show only marginal or no X-ray detection. The presence of Bowen mechanism and especially the high-ionization coronal lines, however, imply the existence of soft X-ray emission, which are likely absorbed in these cases. It is possible that viewing angle affects play an important role, similar to regular TDEs. The X-ray bright BFFs, particularly AT2019aalc and AT2020afhd, exhibit recurrent soft X-ray flares \citep{lancel} (in the case of AT2019aalc followed by the optical QPOs). These were explained \citep{2025arXiv251112477W,2025arXiv250706296J} by precessing accretion disks caused by the Lense-Thirring effect \citep{1918PhyZ...19..156L} in these two cases. The infrared dust echo -- consistent with the detection of strong high-ionization coronal lines in the vast majority of known BFFs (see Figure 13. in \cite{lancel}) -- may be a common feature of these events, which is not surprising given that these transients often occur in AGN, i.e., in dusty environments. Infrared spectroscopy of newly discovered BFFs will be essential to further study the coronal lines in these systems.

Interestingly, at least some BFFs, including AT2019aalc, AT2019avd and J012026, exhibited multi-peaked Balmer lines. In the case of J012026, this feature may be linked to a face-on elliptical accretion disk \citep{j01} similar to the TDE-AGN AT2020zso \citep{Wevers22}. Alternatively, disk precession might also explain the complex Balmer line structures. Namely, asymmetric illumination of the broad line region, can drive time-evolving red-to-blue asymmetries of the broad emission lines \citep{2025arXiv251109626K}.

BFFs may represent the hidden population of TDEs occurring in AGNs. Interactions between the stellar debris and the accretion disk would explain the disk precession and the observational signatures of this phenomena, such as the QPOs and the complex structure of the Balmer lines. The presence of multi-band QPOs due to disk precession were predicted by the TDE-AGN simulations of \cite{2025ApJ...993..244Z}, moreover, other light curve signatures, such as the precursor flare in AT2022fpx also explainable within this TDE-AGN framework. The bumps that are shown in Figure \ref{fig2}. might, in fact, appear due to dips in the optical light curves that may also be related to disk precession: namely to the geometric obscuration by a tilted inner disk in cases of retrograde and perpendicular orbital inclination of the disrupted star \citep{2025ApJ...993..244Z}. However, accretion disk instabilities provide another plausible explanation for AT2019aalc \citep{Marzena} and maybe other BFFs as well. Moreover, a combination of the two scenarios is entirely possible: a star passing through the disk can perturb the pre-existing disk and trigger instabilities \citep{2024MNRAS.527.8103R}.

Systematic spectroscopic and photometric monitoring of future BFFs will be key to understanding the nature of these interesting nuclear transients. Future surveys like the Vera Rubin Observatory Legacy Survey of Space and Time (LSST) and La Silla Schmidt Southern Survey (LS4; \cite{2025PASP..137i4204M}) are expected to reveal a larger population of BFFs.

\section*{Acknowledgments}
PMV acknowledges the support from the DFG via the Collaborative Research Center SFB1491 \textit{Cosmic Interacting Matters - From Source to Signal}.\newline The ZTF forced-photometry service was funded under the Heising-Simons Foundation grant \#12540303 (PI: Graham).\newline This work has made use of data from the Asteroid Terrestrial-impact Last Alert System (ATLAS) project. The Asteroid Terrestrial-impact Last Alert System (ATLAS) project is primarily funded to search for near earth asteroids through NASA grants NN12AR55G, 80NSSC18K0284, and 80NSSC18K1575; byproducts of the NEO search include images and catalogs from the survey area. This work was partially funded by Kepler/K2 grant J1944/80NSSC19K0112 and HST GO-15889, and STFC grants ST/T000198/1 and ST/S006109/1. The ATLAS science products have been made possible through the contributions of the University of Hawaii Institute for Astronomy, the Queen’s University Belfast, the Space Telescope Science Institute, the South African Astronomical Observatory, and The Millennium Institute of Astrophysics (MAS), Chile.

\subsection*{Author contributions}

The author solely performed the observational analysis, theoretical interpretation, figure production, and manuscript writing. 

\subsection*{Financial disclosure}

None reported.

\subsection*{Conflict of interest}

The authors declare no potential conflict of interests.

\bibliography{Wiley-ASNA}

\end{document}